\documentclass{MYws}
\def\refjl#1#2#3#4#5#6{\bibitem{#1} #2, \Journal{#3}{#4}{#6}{#5}.}

\def\EPJC{{\em Eur. Phys. J.} C}

\def\NPPS{\em Nucl. Phys. B (Proc. Suppl.)}

\def\PR{\em Phys. Rev.}

\def\APNY{\em Ann. Phys., NY\/}
\def\NC{\em Nuovo Cimento}

\def\RPP{\em Rep. Prog. Phys.}

\def\PPNP{\em Prog. Part. Nucl. Phys.}

\newcommand{\eqn}[1]{(\ref{#1})}
\newcommand{\no}{\nonumber}
\newcommand{\bel}[1]{\be\label{#1}}
\newcommand{\ba}{\begin{array}{c}}
\newcommand{\bat}{\begin{array}{cc}}
\newcommand{\ea}{\end{array}}
\newcommand{\beqn}{\begin{eqnarray}}
\newcommand{\eeqn}{\end{eqnarray}}

\newcommand{\bi}{\begin{itemize}}
\newcommand{\ei}{\end{itemize}}
\newcommand{\chpt}{$\chi$PT}

\newcommand{\lsim}{~{}_{\textstyle\sim}^{\textstyle <}~}

\def\gap{\;\lower3pt\hbox{$\buildrel > \over \sim$}\;}
\def\lap{\;\lower3pt\hbox{$\buildrel < \over \sim$}\;}

\newcommand{\cL}{{\cal L}}

\newcommand{\cM}{{\cal M}}

\newcommand{\cA}{{\cal A}}

\newcommand{\Tr}{\mbox{\rm Tr}}
\newcommand{\e}{\mbox{\rm e}}

\begin{document}

\title{Colourless Mesons in a Polychromatic World}

\author{A. Pich}

\address{Departament de F\'{\i}sica Te\`orica, IFIC,
Universitat de Val\`encia -- CSIC\\
Edifici d'Instituts de Paterna,
Apt. 22085, E-46071 Val\`encia, Spain, \\   
E-mail: Antonio.Pich@uv.es}


\maketitle

\abstracts{The $N_C\to\infty$ limit of QCD gives a useful
approximation scheme to the physical hadronic world. A brief
overview of the mesonic sector is presented. The large--$N_C$
constraints on the low-energy chiral couplings are summarized
and the role of unitarity corrections is discussed. As an
important illustration of the $1/N_C$ expansion techniques,
the Standard Model prediction of $\varepsilon'/\varepsilon$ 
is reviewed.}

\section{Mesons at Large $\mathbf{N_C}$}
\label{sec:introduction}

The limit of an infinite number of quark colours
turns out to be a very useful starting point to understand many
features of the strong interaction.\cite{HO:74,WI:79}
The $SU(N_C)$ gauge theory simplifies considerably at
$N_C\to\infty$, while keeping the most essential properties of
QCD. Choosing the coupling constant $g_s$ to be of
$O\left(1/\sqrt{N_C}\,\right)$,
{\it i.e.}, taking the large--$N_C$ limit with $\alpha_s N_C$ fixed, there
exists a systematic expansion in powers of $1/N_C$, which for $N_C=3$
provides a good quantitative approximation scheme to the 
hadronic world.\cite{MA:98}
The combinatorics of Feynman diagrams at large $N_C$ results
in simple counting rules, which characterize the $1/N_C$ expansion:
\begin{enumerate}
\item Dominance of planar diagrams with an arbitrary number of gluon exchanges
(and a single quark loop at the edge for matrix elements of quark  bilinears).
\item Non-planar diagrams are suppressed by factors of $1/N_C^2$.
\item Internal quark loops are suppressed by factors of $1/N_C$.
\end{enumerate}

The summation of the leading planar diagrams is a very formidable task,
which we are still unable to perform. Nevertheless, making the very
plausible assumption that colour confinement persists at $N_C\to\infty$,
a very successful 
picture of the meson world emerges.

Let us consider a generic $n$-point function of local quark bilinears\
$J = \bar q\,\Gamma q$:
\bel{eq:n-point}
\langle T\left(J_1 \cdots J_n\right)\rangle \sim O(N_C)\, .
\ee
%
\begin{figure}[thb]
\centerline{\epsfxsize =10.cm \epsfbox{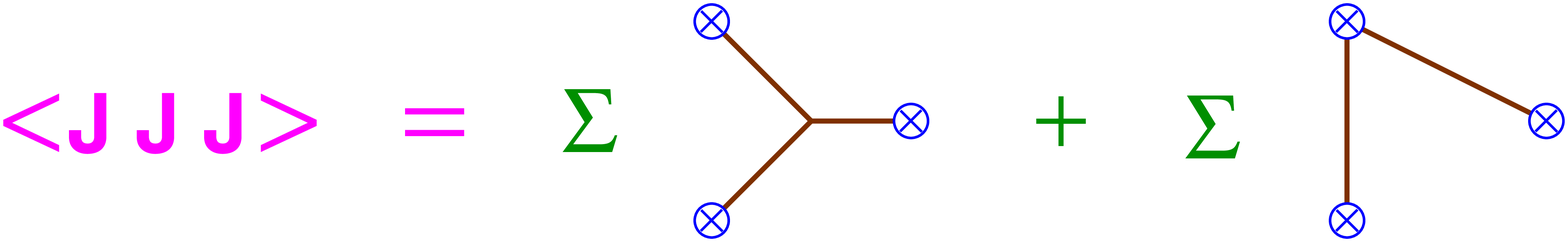}}
\caption{3-point function at large $N_C$.}
\end{figure}
%
A simple diagrammatic analysis shows that 
at large $N_C$
the only
singularities are one-meson poles.\cite{WI:79}
For instance, the two-point function takes the form:
\bel{eq:2-point}
\langle J(k) \, J(-k)\rangle = \sum_n {f_n^2\over k^2-M_n^2} \, .
\ee
Thus:
\begin{itemize}
\item[i)] 
$f_n = \langle 0|J|n\rangle \sim O(\sqrt{N_C}\,)$ \ and \
$M_n\sim O(1)$.

\item[ii)] There are an infinite number of meson states, since 
$\langle J(k) \, J(-k)\rangle$
behaves logarithmically for large $k^2$.

\item[iii)] Mesons are free, stable and non-interacting.
\end{itemize}

At $N_C\to\infty$, the $n$-point functions are given
by sums of tree diagrams with free meson propagators and 
effective local interaction vertices among $m$ mesons,
which scale as \ $V_m\sim O(N_C^{1-m/2})$. Moreover,
$\langle 0|J|M_1\cdots M_m\rangle\sim O(N_C^{1-m/2})$.
Each additional meson coupled to the current $J$ or to
an interaction vertex brings then a suppression factor
$1/\sqrt{N_C}$.

Including gauge-invariant gluon operators, 
such as $J_G = \Tr\left(G_{\mu\nu}G^{\mu\nu}\right)$,
the diagrammatic analysis can be easily extended to glue states.\cite{WI:79}
Since
$\langle T\left(J_{G_1} \cdots J_{G_n}\right)\rangle \sim O(N_C^2)$,
one derives the large--$N_C$ counting rules \
$\langle 0|J_G|G_1\cdots G_m\rangle\sim O(N_C^{2-m})$ \
and \ $V[G_1,\cdots,G_m]\sim O(N_C^{2-m})$.
Thus, at $N_C\to\infty$, glueballs are also free, stable, 
non-interacting and infinite in number. From the mixed correlators 
$\langle T\left(J_1 \cdots J_n J_{G_1} \cdots J_{G_m}\right)\rangle 
\sim O(N_C)$, one gets 
$V[M_1,\cdots,M_p;G_1,\cdots,G_q]\sim O(N_C^{1-q-p/2})$.
Therefore,
glueballs and mesons decouple at large $N_C$, their mixing
being suppressed by a factor $1/\sqrt{N_C}$.

Many known phenomenological features of the hadronic world
are easily understood at lowest order in the $1/N_C$ expansion:
suppression of the $\bar q q$ sea (exotics),
quark model spectroscopy, Zweig's rule, light SU(3) meson nonets,
narrow resonances, multiparticle decays dominated by resonant
two-body final states, etc.
In some cases, the large--$N_C$ limit is in fact the only
known theoretical explanation that is sufficiently general.
Clearly, the expansion in powers of $1/N_C$ appears to be
a sensible physical approximation at $N_C = 3$.

The large--$N_C$ limit provides a weak coupling regime
to perform quantitative QCD studies.
At leading order in $1/N_C$, the scattering amplitudes
are given by sums of tree diagrams with physical
hadrons exchanged. Crossing and unitarity imply that this
sum is the tree approximation to some local effective
Lagrangian. 
Higher-order corrections correspond
to hadronic loop diagrams.

\section{Chiral Symmetry}
\label{sec:chpt}

With $n_f$ massless quark flavours, the QCD Lagrangian \
[$\bar q = (\bar u,\bar d,\ldots)$]
%
\bel{eq:LQCD}
{\cL}_{\mathrm{QCD}}^0 = -{1\over 4}\, G^a_{\mu\nu} G^{\mu\nu}_a
 + i \,\bar q_L^{\phantom{.}} \gamma^\mu D_\mu q_L^{\phantom{.}}  
 + i \,\bar q_R^{\phantom{.}} \gamma^\mu D_\mu q_R^{\phantom{.}}
\ee
is invariant under global \ $U(n_f)_L\otimes U(n_f)_R$
transformations of the left- and right-handed quarks in flavour space:
\raisebox{.15ex}{
$q_{L,R}^{\phantom{.}}\to  g_{L,R}^{\phantom{.}} 
\; q_{L,R}^{\phantom{.}}\,$}, \ \raisebox{.15ex}{
$g_{L,R}^{\phantom{.}} \in U(n_f)_{L,R}^{\phantom{.}}\,$}.
Under very general assumptions it has been shown that, at
$N_C\to\infty$, the symmetry group must spontaneously
break down to the diagonal $U(n_f)_{L+R}$.\cite{CW:80}
According to Goldstone's theorem,\cite{GO:61}
$n_f^2$ pseudoscalar massless bosons appear in the theory,
which for $n_f=3$ can be identified with the $U(3)$ multiplet
\vskip -.1cm
$$
\Phi =  \pmatrix{
 {1\over\sqrt 2}\pi^0 + {1\over\sqrt 6}\eta_8
 + {1\over\sqrt 3}\eta_1 & \pi^+ & K^+ \cr
\pi^- & - {1\over\sqrt 2}\pi^0 + {1\over\sqrt 6}\eta_8 
 + {1\over\sqrt 3}\eta_1  & K^0 \cr 
 K^- & \bar{K}^0 & - {2 \over\sqrt 6}\eta_8 
 + {1\over\sqrt 3}\eta_1  & }.
$$
\vskip .1cm\noindent
The unitary matrix
\bel{eq:u_parametrization}
U(\phi) =  u(\phi)^2 =
\exp{\left\{i\sqrt{2}\Phi/f\right\}}
\ee
gives a very convenient parameterization of the
Goldstone fields. Under the chiral group it transforms
as \ $U(\phi)\to g_R^{\phantom{.}}\, U(\phi)\, g_L^\dagger$.

The Goldstone nature of the pseudoscalar mesons implies strong
constraints on their interactions, which can be most easily analyzed
on the basis of an effective Lagrangian.\cite{WE:79}
Since there is a mass gap separating the pseudoscalar nonet from the
rest of the hadronic spectrum, we can build an effective field 
theory\cite{EFT} (EFT) containing only the Goldstone modes.
Moreover, the low-energy effective Lagrangian can be organized in
terms of increasing powers of momenta (derivatives).

Let us consider an extended QCD Lagrangian, with quark
couplings to external Hermitian matrix-valued fields
$l_\mu$, $r_\mu$, $s$, $p$\/ :
\be
\cL_{\mathrm{QCD}} = \cL^0_{\mathrm{QCD}} +
\bar q_L^{\phantom{.}} \gamma^\mu l_\mu\, q_L^{\phantom{.}} 
+ q_R^{\phantom{.}} \gamma^\mu r_\mu\, q_R^{\phantom{.}} -
\bar q_L^{\phantom{.}} (s - i p)\, q_R^{\phantom{.}} -  
\bar q_R^{\phantom{.}} (s + i p)\, q_L^{\phantom{.}}\, .
\label{eq:extendedqcd}
\ee
The external fields can be used to incorporate the
electromagnetic and semileptonic weak interactions, and the
explicit breaking of chiral symmetry through the quark masses:
\bel{eq:breaking}
s = \cM + \ldots\, , \qquad\qquad \cM= \hbox{\rm diag}(m_u,m_d,m_s)
\, .
\ee
At lowest order in derivatives and quark masses,
the most general effective Lagrangian
consistent with chiral symmetry has the form:\cite{GL:85}
\bel{eq:lowestorder}
\cL_2 = {f^2\over 4}\,
\langle D_\mu U^\dagger D^\mu U \, + \, U^\dagger\chi  \, 
+  \,\chi^\dagger U\rangle\, ,
\qquad\qquad
\chi \equiv 2 B_0 \, (s + i p) \, ,
\ee
where \
$D_\mu U = \partial_\mu U -i r_\mu U + i U\, l_\mu$ ,
$\langle A\rangle$ denotes the flavour trace of the matrix $A$
and $B_0$ is a constant, which, like $f$, is not fixed by
symmetry requirements alone.
Taking functional derivatives with respect to the appropriate
external fields, one finds that $f$ equals the pion decay constant
(at lowest order) $f = f_\pi = 92.4$~MeV, while $B_0$ is
related to the quark condensate:
\bel{eq:B0}
B_0 = -{\langle\bar q q \rangle\over f^2} =
{M_\pi^2\over m_u + m_d} = {M_{K^0}^2\over m_s + m_d} =
{M_{K^\pm}^2\over m_s + m_u}\, .
\ee

Formally, the chiral Lagrangian could be computed 
(non-perturbatively) from the QCD generating functional.
The leading-order terms in $1/N_C$ should be of $O(N_C)$, like the
corresponding correlation functions of fermion bilinears.
Moreover, they should have a single flavour trace since
diagrams with $n$ quark loops have $n$ flavour traces and are
of \ $O(N_C^{2-n})$. The Lagrangian $\cL_2$ obeys the correct
$N_C$ counting rules:
$f^2\sim O(N_C)$, $B_0\sim M_\phi^2\sim U(\phi)\sim O(1)$.
The $U(\phi)$ matrix generates an expansion in powers
of $\phi/f$, giving the required $1/\sqrt{N_C}$ suppression
for each additional meson field. Clearly, interaction
vertices with $n$ mesons scale as 
$V_n\sim f^{2-n}\sim O(N_C^{1-n/2})$.
Since $\cL_2$ has an overall factor of $N_C$ and $U$ is
$N_C$-independent, the $1/N_C$ expansion is equivalent to a 
semiclassical expansion. Quantum corrections computed with the
chiral Lagrangian will have a $1/N_C$ suppression for each loop.

At $O(p^4)$, the conventional $SU(3)_L\otimes SU(3)_R$-invariant
chiral Lagrangian is usually written as:\cite{GL:85}
\beqn\label{eq:l4}
\cL_4 & = &
L_1 \,\langle D_\mu U^\dagger D^\mu U\rangle^2 \, + \,
L_2 \,\langle D_\mu U^\dagger D_\nu U\rangle\,
   \langle D^\mu U^\dagger D^\nu U\rangle
\no\\ &&\mbox{}
+ L_3 \,\langle D_\mu U^\dagger D^\mu U D_\nu U^\dagger
D^\nu U\rangle\,
+ \, L_4 \,\langle D_\mu U^\dagger D^\mu U\rangle\,
   \langle U^\dagger\chi +  \chi^\dagger U \rangle
\no\\ &&\mbox{}
+ L_5 \,\langle D_\mu U^\dagger D^\mu U \left( U^\dagger\chi +
\chi^\dagger U \right)\rangle\,
+ \, L_6 \,\langle U^\dagger\chi +  \chi^\dagger U \rangle^2
\\ &&\mbox{}
+ L_7 \,\langle U^\dagger\chi -  \chi^\dagger U \rangle^2\,
+ \, L_8 \,\langle\chi^\dagger U \chi^\dagger U
+ U^\dagger\chi U^\dagger\chi\rangle
\no \\ &&\mbox{}
- i L_9 \,\langle F_R^{\mu\nu} D_\mu U D_\nu U^\dagger +
     F_L^{\mu\nu} D_\mu U^\dagger D_\nu U\rangle\,
+ \, L_{10} \,\langle U^\dagger F_R^{\mu\nu} U F_{L\mu\nu} \rangle
 \, ,
\no
\eeqn
where $F^{\mu\nu}_{L,R}$ are field-strength tensors of the
$l^\mu$ and $r^\mu$ flavour fields.

{\renewcommand{\arraystretch}{1.2} 
 \begin{table}[tbh]
\label{tab:Lcouplings}
\begin{center}
\footnotesize
\begin{tabular}{|c|c|c|c|c|c|}
\hline
$i$ & \raisebox{0pt}[12pt][7pt]{$L_i^r(M_\rho)$} & $O(N_C)$  & Source 
& $\Gamma_i$ & $L_{i}^{N_C\to\infty}$
\\
\hline
\raisebox{0pt}[12pt]{$2L_1-L_2$} & $-0.6\pm0.6$ & $O(1)$ &
$K_{e4}$, $\pi\pi\to\pi\pi$ & $3/16$ & 0
\\
$L_2$ & $\hphantom{-}1.4\pm0.3$ & $O(N_C)$ &
$K_{e4}$, $\pi\pi\to\pi\pi$ & $3/16$ & $\phantom{-}1.8$
\\
$L_3$ & $-3.5\pm1.1$ & $O(N_C)$ & $K_{e4}$, $\pi\pi\to\pi\pi$
& 0 & $-4.3$
\\
$L_4$ & $-0.3\pm0.5$ & $O(1)$ & Zweig rule & $1/8$ & 0
\\
$L_5$ & $\hphantom{-}1.4\pm0.5$ & $O(N_C)$ & $F_K : F_\pi$
& $3/8$ & $\phantom{-}2.1$
\\
$L_6$ & $-0.2\pm0.3$ & $O(1)$ & Zweig rule & $11/144$ & 0
\\
$L_7$ & $-0.4\pm0.2$ & $O(1)$ & GMO, $L_5$, $L_8$ & 0 & $-0.3$
\\
$L_8$ & $\hphantom{-}0.9\pm0.3$ & $O(N_C)$ & $M_\phi$, $L_5$
& $5/48$ & $\phantom{-}0.8$
\\
$L_9$ & $\hphantom{-}6.9\pm0.7$ & $O(N_C)$ &
$\langle r^2\rangle^\pi_V$ & $1/4$ & $\phantom{-}7.1$
\\
$L_{10}$ & $-5.5\pm0.7$ & $O(N_C)$ & $\pi\to e\nu\gamma$
& $-1/4$ & $-5.4$
\\[3pt] \hline
\end{tabular}
\end{center}
\caption{Phenomenological values of the
renormalized couplings $L_i^r(M_\rho)$ in units of $10^{-3}$.
The fourth column shows the source used to get this information.
The large--$N_C$ predictions obtained within the single-resonance 
approximation are given in the last column.}
\end{table}}

Thus, at $O(p^4)$ we need ten additional coupling constants
$L_i$
to determine the low-energy behaviour of the Green functions.
Terms with a single trace are of $O(N_C)$, while those
with two traces should be of $O(1)$. However, a
$3\times 3$ matrix relation has been used to eliminate the
additional structure
$c\,\langle D_\mu U^\dagger D_\nu U D^\mu U^\dagger
D^\nu U\rangle$ with the result
$2\delta L_1 = \delta L_2 = -\frac{1}{2}\delta L_3 = c \sim O(N_C)$.
As shown in Table~\ref{tab:Lcouplings}, 
the phenomenologically determined values\cite{EC:95,PI:95}
of those couplings follow the pattern suggested by the $1/N_C$
counting rules.
Moreover, their average order of magnitude, 
$L_i \sim f^2/(4\Lambda_\chi^2)\sim 2\times 10^{-3}$,
suggests a chiral
symmetry-breaking scale $\Lambda_\chi\sim 1$~GeV.

One-loop graphs with the lowest-order Lagrangian $\cL_2$ contribute
also at $O(p^4)$ in the chiral expansion, but they are
suppressed by a factor of $1/N_C$. 
Their divergent parts are renormalized by the
$\cL_4$ couplings:
\bel{eq:renormalization}
L_i\, =\, L_i^r(\mu)\, + \,\Gamma_i\, {\mu^{D-4}\over 32 \pi^2}\, \left\{
{2\over D-4} + \gamma_E - \log{(4\pi)} - 1 \right\}  .
\ee
This introduces a renormalization scale dependence,
\bel{eq:l_running}
L_i^r(\mu_2)  =  L_i^r(\mu_1)  +  {\Gamma_i\over (4\pi)^2}
\,\log{\left({\mu_1\over\mu_2}\right)} ,
\ee
which is subleading in $1/N_C$.  
The phenomenological couplings given in Table~\ref{tab:Lcouplings}
have been normalized at $\mu=M_\rho$.

The chiral loops generate non-polynomial contributions,
with logarithms and threshold factors as required by unitarity,
which are completely predicted as functions of $f$ and the Goldstone
masses.
Although they are suppressed by a factor of $1/N_C$, the chiral logarithms
can be numerically important since 
$\frac{1}{N_C}\log{(\Lambda^2_\chi/M^2_\pi)}\sim 4/3$.

\subsection{Anomalies}

Since chiral symmetry is explicitly violated by fermion 
anomalies at the fundamental QCD level,\cite{AD:69}
we need to add a functional $Z_{\cA}$ with the property that its
change under chiral transformations reproduces
the anomalous change of the QCD generating functional.
For the non-Abelian anomalies associated with the external
sources $l_\mu$ and $r_\mu$, such a functional 
was first constructed by Wess and Zumino,\cite{WZ:71}
and reformulated in a nice geometrical way by Witten.\cite{WI:83}
It is an $O(p^4)$ effect, which is completely calculable with
no free parameters. This contribution is of \ $O(N_C)$,
because it is generated by a triangle quark loop
coupled to external sources.

Much more subtle is the $U(1)_A$ gluonic anomaly which breaks
the conservation of the singlet axial quark current in the
chiral limit:
\bel{eq:omega}
\partial_\mu\left(\bar q\gamma^\mu\gamma_5 q\right) =
2\, n_f \omega
\qquad ; \qquad
\omega =
{\alpha_s\over 16\pi}\, \epsilon^{\mu\nu\rho\sigma}\, 
G_{\mu\nu} G_{\rho\sigma} \, .
\ee
The corresponding anomalous change of the QCD generating functional 
can be accounted for by adding a term
$\Delta\cL_{\mathrm{QCD}} = -\theta\,\omega$ 
with the appropriate chiral transformation for the so-called 
vacuum angle $\theta(x)$.\cite{KL:00}
Notice that in the large--$N_C$ limit the $U(1)_A$ anomaly is 
absent.\cite{WI:79b}

To lowest non-trivial order in $1/N_C$, the chiral symmetry breaking
effect induced by the $U(1)_A$ anomaly can be taken into
account in the effective low-energy theory, through the 
term\cite{DVV:80}      
\bel{eq:anom_term}
\cL_{U(1)_A} \, = \, - {f^2 \over 4} {a \over N_C} \, 
\left\{\theta - {i\over 2} \left[\log{(\det{U})} - 
\log{(\det{U^\dagger})}\right] \right\}^2  ,
\ee
which breaks $U(3)_L \otimes U(3)_R$ but preserves
$SU(3)_L \otimes SU(3)_R \otimes U(1)_V$. 

The parameter $\, a \,$ has dimensions of mass squared and, 
with the factor $1/N_C$ pulled out, 
is booked to be of $O(1)$ in the large--$N_C$
counting rules. Its value is not fixed by symmetry requirements alone;
it depends crucially on the dynamics of instantons. In the presence
of the term \eqn{eq:anom_term}, 
the $\eta_1$ field becomes massive even in the
chiral limit:
\bel{eq:M_eta1}
M_{\eta_1}^2 = 3\, {a\over N_C} + O(\cM) \, .
\ee

Owing to the large mass of the $\eta'$, the effect of the $U(1)_A$ anomaly
cannot be treated as a small perturbation. Rather, one should keep
the term \eqn{eq:anom_term} together with the lowest-order Lagrangian
\eqn{eq:lowestorder}. It is possible to build a consistent combined 
expansion
in powers of momenta, quark masses and $1/N_C$, by counting the
relative magnitude of these parameters as:\cite{LE:96}
\bel{eq:U(3)_counting}
O(p^2) \sim O(\cM) \sim O(1/N_C) \, .
\ee
This expansion has been already analyzed at
the next-to-leading order.\cite{KL:00,HLPT:97,KA:02}

\section{Resonance Chiral Theory}

Let us consider a chiral-invariant Lagrangian
$\cL(U,V,A,S,P)$, describing the couplings of resonance nonet multiplets
of the type $V(1^{--})$, $A(1^{++})$, $S(0^{++})$ and $P(0^{-+})$ to
the Goldstone bosons:\cite{EGPR:89}
\beqn\label{eq:R_int}
\cL_2[V(1^{--})] &\; =\; & \sum_i\;\left\{ {F_{V_i}\over 2\sqrt{2}}\;
   \langle V_i^{\mu\nu} f_{+ \, \mu\nu}\rangle\, +\,
   {i\, G_{V_i}\over \sqrt{2}} \,\,\langle V_i^{\mu\nu} u_\mu u_\nu\rangle
   \right\}\, , 
\no\\
\cL_2[A(1^{++})] & = & \sum_i\; {F_{A_i}\over 2\sqrt{2}} \;
   \langle A_i^{\mu\nu} f_{- \, \mu\nu} \rangle\, ,
\\
\cL_2[S(0^{++})] & = & \sum_i\;\biggl\{ c_{d_i} \; \langle S_i\, u^\mu
u_\mu\rangle\, +\, c_{m_i} \; \langle S_i\, \chi_+ \rangle\biggr\}\, ,
\no\\
\cL_2[P(0^{-+})] &=&\sum_i\; i\, d_{m_i}\;\langle P_i\, \chi_- \rangle\, ,
\no
\eeqn
where \
$u_\mu \equiv i\, u^\dagger D_\mu U u^\dagger$, \
$f^{\mu\nu}_\pm\equiv u F_L^{\mu\nu} u^\dagger\pm  u^\dagger F_R^{\mu\nu} u$
\ and \
$\chi_\pm\equiv u^\dagger\chi u^\dagger\pm u\chi^\dagger u$.
The resonance couplings 
$F_{V_i}$, $G_{V_i}$, $F_{A_i}$, $c_{d_i}$, $c_{m_i}$ and $d_{m_i}$
are of \ $O\left(\sqrt{N_C}\,\right)$.

The lightest resonances have an important impact on the
low-energy dynamics of the pseudoscalar bosons.
Below the resonance mass scale, the singularity associated with the
pole of a resonance propagator is replaced by the corresponding
momentum expansion; therefore, the exchange of virtual resonances generates
derivative Goldstone couplings proportional to powers of $1/M_R^2$.
At lowest order in derivatives, this gives the large--$N_C$ predictions
for the $O(p^4)$ couplings of chiral perturbation theory
(\chpt):\cite{EGPR:89}
\beqn\label{eq:vmd_results}
2\, L_1 = L_2 = \sum_i\; {G_{V_i}^2\over 4\, M_{V_i}^2}\, , & \qquad &
L_3 = \sum_i\;\left\{ -{3\, G_{V_i}^2\over 4\, M_{V_i}^2} +
{c_{d_i}^2\over 2\, M_{S_i}^2}\right\} \, ,
\no\\
L_5 = \sum_i\; {c_{d_i}\, c_{m_i}\over M_{S_i}^2} \, ,\hskip .89cm &&
L_8 = \sum_i\;\left\{ {c_{m_i}^2\over 2\, M_{S_i}^2} -
{d_{m_i}^2\over 2\, M_{P_i}^2}\right\} \, ,
\\
L_9 = \sum_i\; {F_{V_i}\, G_{V_i}\over 2\, M_{V_i}^2}\, ,\hskip .77cm &&
L_{10} = \sum_i\;\left\{ {F_{A_i}^2\over 4\, M_{A_i}^2}
 - {F_{V_i}^2\over 4\, M_{V_i}^2}\right\}  \, .
\no\eeqn
All these couplings are of $O(N_C)$, in agreement with the
counting made in Table~\ref{tab:Lcouplings}, while for the 
couplings of $O(1)$ we get
$2\, L_1-L_2 = L_4 = L_6 = L_7 = 0$.

Owing to the $U(1)_A$ anomaly, the $\eta_1$ field is massive and it is often
integrated out from the low-energy chiral theory. In that case,
the $SU(3)_L\otimes SU(3)_R$ chiral coupling $L_7$ gets a contribution
from $\eta_1$ exchange:\cite{GL:85,EGPR:89}
\bel{eq:L7}
L_7 = - {\tilde{d}_m^2\over 2\, M^2_{\eta_1}} \, ,
\qquad\qquad\qquad
\tilde{d}_m = -{f\over\sqrt{24}} \, .
\ee
Since, $M^2_{\eta_1}\sim O\left(1/N_C,\cM\right)$, 
the coupling $L_7$ could then\cite{GL:85} be considered
of $O(N_C^2)$.
However, the large--$N_C$ counting is no longer consistent if one
takes the limit of a heavy $\eta_1$ mass
($N_C$ small) while keeping $m_s$ small.\cite{PdR:95}

\subsection{Short-Distance Constraints}

The short-distance properties of the underlying QCD dynamics
impose some constraints on the low-energy EFT parameters:\cite{EGLPR:89}

\begin{enumerate}

\item {\it Vector Form Factor.} \
At leading order in $1/N_C$, the two-Goldstone matrix element of the
vector current, is characterized by
\bel{eq:VFF}
F_V(t)\, =\, 1\, + \, \sum_i\,
{F_{V_i}\, G_{V_i}\over f^2}\; {t\over M_{V_i}^2-t} \, .
\ee
Since the vector form factor $F_V(t)$ should vanish at infinite momentum 
transfer $t$, the resonance couplings should satisfy
\bel{eq:SD1}
\sum_i\, F_{V_i}\, G_{V_i}\, =\, f^2\, .
\ee

\item {\it Axial Form Factor.} \
The matrix element of the axial current between one Goldstone and
one photon is parameterized by the axial form factor. From the resonance
Lagrangian \eqn{eq:R_int}, one gets
\bel{eq:AFF}
G_A(t)\, =\, \sum_i\, \left\{
{2\, F_{V_i}\, G_{V_i}- F_{V_i}^2\over M_{V_i}^2}\, +\,
{F_{A_i}^2\over M_{A_i}^2-t} \right\}\, ,
\ee
which vanishes at $t\to\infty$ provided that
\bel{eq:SD2}
\sum_i\,
{2\, F_{V_i}\, G_{V_i}- F_{V_i}^2\over M_{V_i}^2}
\, =\, 0\, .
\ee

\item {\it Weinberg Sum Rules.} \
The two-point function built from a left-handed and a right-handed
vector quark current defines the correlator
\bel{eq:WSR}
\Pi_{LR}(t)\, =\, {f^2\over t} \, +\,\sum_i\,
{F_{V_i}^2\over M_{V_i}^2-t} \, -\,\sum_i\,
{F_{A_i}^2\over M_{A_i}^2-t} \, .
\ee
Since gluonic interactions preserve chirality,
$\Pi_{LR}(t)$ satifies an unsubtracted dispersion relation.
Moreover,\cite{FNdR:79} 
in the chiral limit it vanishes faster than $1/t^2$
when $t\to\infty$. This implies the well-known conditions:\cite{WE:67}
\bel{eq:SD3}
\sum_i\,\left( F_{V_i}^2 - F_{A_i}^2\right) = f^2 \, ,
\qquad\qquad
\sum_i\,\left( M_{V_i}^2 F_{V_i}^2 - M_{A_i}^2 F_{A_i}^2\right) 
= 0 \, .
\ee
The second relation is correct up to very small quark-mass contributions.

\item {\it Scalar Form Factor.} \
The two-pseudoscalar matrix element of the scalar quark current
contains another dynamical form factor, which for the $K\pi$
case takes the form:\cite{JOP:02}
\bel{eq:SFF}
F^S_{K\pi}(t)\, =\, 1\, + \, \sum_i\,
{4\, c_{m_i}\over f^2}\left\{ c_{d_i} +
\left( c_{m_i}-c_{d_i}\right)\,
{M_K^2-M_\pi^2\over M_{S_i}^2}\right\}
 {t\over M_{S_i}^2-t} \, .
\ee
Requiring $F^S(t)$ to vanish at $t\to\infty$, one gets the
constraints:\cite{JOP:02}
\bel{eq:SD4}
4\,\sum_i\,c_{d_i}\, c_{m_i} = f^2 \, ,
\qquad\qquad
\sum_i\,  {c_{m_i}\over M_{S_i}^2}\left( c_{m_i}-c_{d_i}\right) = 0 \, .
\ee

\item {\it $SS-PP$ Sum Rules.} \
The two-point correlation functions of two scalar or two pseudoscalar
currents would be equal if chirality was absolutely preserved. Their
difference is easily computed in the hadronic EFT:
\bel{eq:SSR}
\Pi_{SS-PP}(t)\, =\, 16\, B_0^2\,\left\{ 
\sum_i\, {c_{m_i}^2\over M_{S_i}^2-t} \, -\,\sum_i\,
{d_{m_i}^2\over M_{P_i}^2-t} \, +\, {f^2\over 8\, t}\right\}\, .
\ee
For massless quarks, $\Pi_{SS-PP}(t)$ vanishes as $1/t^2$ when
$t\to\infty$, with a coefficient proportional to\cite{SVZ:79}
$\alpha_s\,\langle\bar q\Gamma q\,\bar q\Gamma q\rangle$.
The vacuum four-quark condensate provides a non-perturbative breaking
of chiral symmetry. In the large--$N_C$ limit, it factorizes as
$\alpha_s\,\langle\bar q q\rangle^2 \sim \alpha_s\, B_0^2$.
Imposing this behaviour on \eqn{eq:SSR}, one gets:\cite{GP:00}
\bel{eq:SD5}
8\,\sum_i\left( c_{m_i}^2 - d_{m_i}^2\right) = f^2  ,
\qquad
\sum_i\left( c_{m_i}^2 M_{S_i}^2 - d_{m_i}^2 M_{P_i}^2\right) = 
{3\,\pi\alpha_s\over 4}\, f^4  .
\ee

\end{enumerate}

\subsection{Single-Resonance Approximation}

Let us approximate each infinite resonance sum 
with the contribution from the first meson nonet
with the given quantum numbers. This is meaningful at
low energies where the contributions from higher-mass states are
suppressed by their corresponding propagators. The
single-resonance approximation (SRA) corresponds to work with a
low-energy EFT below the scale of the second resonance multiplets.
The resulting short-distance constraints are nothing else than the
matching conditions between this EFT and the underlying QCD dynamics.
Thus, we are assuming that the
short-distance operator product expansion provides an acceptable
description at energies above 1.5~GeV.

Within the SRA, Eqs.~\eqn{eq:SD1}, \eqn{eq:SD2} and \eqn{eq:SD3}
determine the vector and axial-vector couplings in terms of $M_V$
and $f$:\cite{EGLPR:89}
\bel{eq:VA_coup}
F_V = 2\, G_V = \sqrt{2}\, F_A = \sqrt{2}\, f \, ,
\qquad\qquad
M_A = \sqrt{2}\, M_V \, .
\ee
The scalar\cite{JOP:02}
and pseudoscalar parameters are obtained 
from \eqn{eq:SD4} and \eqn{eq:SD5}:
\bel{eq:SP_coup}
c_m = c_d = \sqrt{2}\, d_m = f/2 \, ,
\qquad\qquad
M_P = \sqrt{2}\, M_S \, \left(1 - \delta\right)^{1/2}\, .
\ee
The last relation involves a small correction \
$\delta \approx 3\,\pi\alpha_s f^2/M_S^2 \sim 0.08\,\alpha_s$,
which we can neglect together with the tiny effects from
light quark masses.

Inserting these predictions into Eqs.~\eqn{eq:vmd_results},
one finally gets all $O(N_C\, p^4)$ \chpt\ couplings, in terms
of $M_V$, $M_S$ and $f$:
\bel{eq:Li_SRA_1}
2\, L_1 = L_2 = \frac{1}{4}\, L_9 = -\frac{1}{3}\, L_{10}
= {f^2\over 8\, M_V^2}\, ,
\ee
\bel{eq:Li_SRA_2}
L_3 = -{3\, f^2\over 8\, M_V^2} + {f^2\over 8\, M_S^2}\, ,
\qquad\quad
L_5 ={f^2\over 4\, M_S^2}\, ,
\qquad\quad
L_8 = {3\, f^2\over 32\, M_S^2}\, .
\ee
The last column in Table~\ref{tab:Lcouplings} shows the
results obtained with $M_V = 0.77$~GeV,
$M_S = 1.0$~GeV and $f=92$~MeV. Also shown is the $L_7$
prediction in \eqn{eq:L7}, taking
$M_{\eta_1} = 0.80$~GeV. The agreement with
the measured values is a clear
success of the large--$N_C$ approximation.
It demonstrates that the lightest resonance multiplets
give indeed the dominant effects at low energies.

The study of other Green functions provides further matching
conditions between the hadronic and fundamental QCD descriptions.
Clearly, it is not possible to satisfy all of them
within the SRA. A useful generalization is the
so-called {\it Minimal Hadronic Ansatz}, which consists of
keeping the minimum number of resonances compatible with all known
short-distance constraints for the problem at hand.\cite{KPdR}
Some $O(p^6)$ \chpt\ couplings have been already
analyzed in this way, by studying an appropriate set of three-point
functions.\cite{KN:01}

\section{Unitarity Corrections}

The \chpt\ loops incorporate the unitarity field theory constraints 
in a pertur\-bative way, order by order in the chiral expansion.
Although subleading in the $1/N_C$ counting, these corrections
may be enhanced by infrared logarithms.
Their effect appears to be crucial for a correct understanding of
some observables, in particular in the scalar sector, because the
S--wave rescattering of two pseudoscalars is very strong.
The combined constraints of analyticity and unitarity make possible
to perform appropriate resummations of chiral logarithms,
which describe the leading $1/N_C$ corrections in the resonance region.

A simple example is provided by the Omn\`es\cite{OM:58} exponentiation
of the pion form factor:\cite{GP:97}
\bel{eq:OmVFF}
F_V(t)\, =\, {M_V^2\over M_V^2 -t}\;\exp{\left\{ 
-{t\over 96\,\pi^2 f^2}\; A^{(\pi)}(t)\right\} }\, ,
\ee
where \ \ [$\sigma_\pi\equiv\sqrt{1-4\, M_\pi^2/t}$]
\bel{eq:Afun}
A^{(\pi)}(t)
\,\equiv\, \sigma_\pi^3\,\log{\left(
{\sigma_\pi + 1\over \sigma_\pi - 1}\right)}\, +\,
\log{\left({M_\pi^2\over \mu^2}\right)}\, +\, 8\, {M_\pi^2\over t}\,
-\, {5\over 3}\, -\, \delta L_9^r(\mu)\, ,
\ee
is the regularized one-loop function describing two intermediate pions
(the small $K\bar K$ loop contribution has been neglected),
which arises here from an integration over the $I=J=1$ \
$\pi\pi$ phase shift at leading order in \chpt,
\bel{eq:omnes}
F_V(t) \, =\, Q_n(t)\,\exp{\left\{\frac{s^n}{\pi} \int^{\infty}_{4M^2_\pi} \, 
\frac{dz}{z^n} \, \frac{\delta^1_1(z)}{z-t-i\epsilon}\right\}} \, . 
\ee
This expression is valid in the elastic region and
has a polynomic ambiguity which is compensated by the
subtraction function $Q_n(t)$. Only the logarithmic
corrections are unambiguous. The ambiguity has been solved by
matching the Omn\`es solution both to the \chpt\ and large--$N_C$
(SRA) results. There remains a local indetermination at higher
orders, made explicit through the constant
$\delta L_9^r(\mu)\equiv 128\,\pi^2\, [ 
L_9^r(\mu) - L_9^{N_C\to\infty}]$, which
is next-to-leading in $1/N_C$ and does not contain any large
infrared logarithm when $\mu\sim M_V$.

\begin{figure}
\begin{center}
\includegraphics[width=\textwidth]{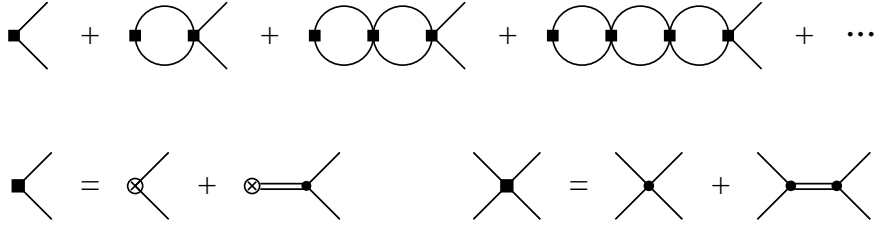}
\end{center}
\caption{Dyson-Schwinger resummation of $F_V(t)$ with
effective vertices.}
\end{figure}

Equation~\eqn{eq:OmVFF} has obvious shortcomings. 
We have used an $O(p^2)$ approximation to the $\pi\pi$ phase shift,
$\delta^1_1(t) = t\, \sigma_\pi^3/(96\,\pi f_\pi^2)$,
which is a very poor (and even wrong) description at the higher end
of the dispersive integration region.
Nevertheless, one can always take a sufficient
number of subtractions to emphasize numerically the low-energy region.
Since our matching has fixed an infinite number of subtractions, this
result should give a good approximation for
values of $s$ not too large. Moreover, this can be phenomenologically
improved with the use of the measured phase shifts.\cite{GPP:00}

A more important question concerns the $\rho$ meson pole, which needs
a proper treatment if one aims to describe physics around or above
the resonance peak. The pole is regulated by the
$\rho$ width, which vanishes at $N_C\to\infty$. The dressed
propagator can be calculated through a Dyson-Schwinger resummation
constructed from effective Goldstone vertices containing both the
local \chpt\ interaction and the resonance-exchange 
contributions:\cite{GP:97,GPP:00,CP:02}
\bel{eq:DSprop}
F_V(t)\, =\, {M_V^2\over M_V^2 -t + \xi_\rho(t) - i\, M_V\,\Gamma_\rho(t)}
\, ,
\ee
where
\bel{eq:xi}
\xi_\rho(t) - i\, M_V\,\Gamma_\rho(t)\, = \,
{t\, M_V^2\over 96\,\pi^2\, f^2}\; A^{(\pi)}(t)\, .
\ee
Thus,
\bel{eq:V_width}
\Gamma_\rho(t)\, = \,\theta (t-4\, M_\pi^2)\;
{t\, M_V\over 96\,\pi\, f^2}\;\sigma_\pi^3\, ,
\ee
which at $t=M_\rho^2$ gives $\Gamma_\rho(M_\rho^2) = 144$~MeV,
in reasonable agreement with the measured $\rho$ width.
The intermediate $K\bar K$ contributions can be included
through a coupled-channel resummation;\cite{CP:02} the only 
modification is the change
$A^{(\pi)}(t)\to A^{(\pi)}(t) +\frac{1}{2}\, A^{(K)}(t)$.

\begin{figure}[thb]
\centerline{
\includegraphics[angle=-90,width=8.25cm]{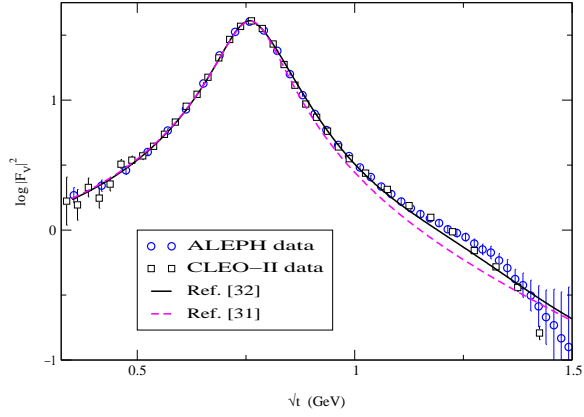}}
\caption{Comparison of $\tau\to\nu_\tau\pi\pi$ data with
Omn\`es predictions for $F_V(t)$.
The dashed line\protect\cite{GP:97} corresponds to Eq.~\protect\eqn{eq:OmVFF},
with the term $i\, M_V\Gamma_\rho(t)$ shifted to the
$\rho$ propagator to regulate the pole. The continuous line is
the 3-subtracted result,\protect\cite{GPP:00} using the
full phase shift~\eqn{eq:PS} for $M_\rho\leq \sqrt{t}$ \
and $\delta^1_1(t)$ data at higher values of $t$.
}
\end{figure}

Equations~\eqn{eq:OmVFF} and \eqn{eq:DSprop} represent different resummations
of higher-order corrections. They agree, by construction, at $O(p^4)$
in \chpt\ and at the leading order in $1/N_C$. 
The result can be further improved
by inserting into the Omn\`es exponential \eqn{eq:omnes}
the phase shift predicted in \eqn{eq:DSprop},
\bel{eq:PS}
\delta^1_1(t)\, =\, \arctan{\left\{ 
{M_V\Gamma_\rho(t)\over M_V^2 -t +\xi_\rho(t)}\right\}}
\,=\, {t\, \sigma_\pi^3\over 96\,\pi f_\pi^2}\, +\, \cdots\, ,
\ee
and imposing the appropriate matching conditions.\cite{GPP:00}

Similar unitarization procedures
have been applied to amplitudes with $I=J=0$, 
which get large corrections from infrared 
chiral logarithms.\cite{JOP:02,oller,PaP:01}

\section{Strangeness-Changing Non-Leptonic Weak Transitions}

\begin{figure}[hbt]
\begin{center}
\includegraphics[width=11.5cm]{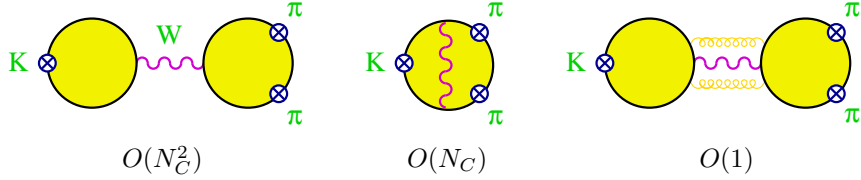}

\mbox{}\hskip .1cm $O(N_C^2)$ \hskip 2.6cm $O(N_C)$
\hskip 2.7cm $O(1)$
\end{center}
\caption{Diagrammatic topologies contributing to $K\to\pi\pi$.}
\end{figure}

Since weak currents factorize at large--$N_C$,
a naive $1/N_C$ description of $K\to\pi\pi$ would imply
$A(K^0\to\pi^0\pi^0)=0$. In terms of isospin amplitudes,
$A_0 \! =\sqrt{2}\, A_2$; {\it i.e.}, there is no $\Delta I=1/2$
enhancement at leading order in $1/N_C$.

{\renewcommand{\arraystretch}{1.0}
\begin{figure}[tbh]    
\setlength{\unitlength}{0.5mm} \centering
\begin{picture}(165,120)
\put(0,0){\makebox(165,120){}}
\thicklines
\put(10,105){\makebox(40,15){Energy Scale}}
\put(58,105){\makebox(36,15){Fields}}
\put(110,105){\makebox(40,15){Effective Theory}}
\put(8,108){\line(1,0){149}}
\put(10,75){\makebox(40,30){$M_Z$}}
\put(58,75){\framebox(38,28){$\ba W, Z, \gamma, g \\       
     \tau, \mu, e, \nu_i \\ t, b, c, s, d, u \ea $}}
\put(110,75){\makebox(40,27){Standard Model}}

\put(10,40){\makebox(40,18){$\lsim m_c$}}
\put(58,40){\framebox(38,18){$\ba  \gamma, g  \, ; \mu ,  e, \nu_i
             \\ s, d, u \ea $}}
\put(110,40){\makebox(40,18){$\cL_{\mathrm{QCD}}^{(n_f=3)}$, \
             $\cL_{\mathrm{eff}}^{\Delta S=1,2}$}}

\put(10,5){\makebox(40,18){$M_K$}}
\put(58,5){\framebox(38,18){$\ba\gamma \; ;\; \mu , e, \nu_i  \\
            \pi, K,\eta  \ea $}}
\put(110,5){\makebox(40,18){$\chi$PT}}
\linethickness{0.3mm}
\put(76,37){\vector(0,-1){11}}
\put(76,72){\vector(0,-1){11}}
\put(80,64.5){OPE}
\put(80,29.5){$N_C\to\infty$}
\end{picture}
\caption[0]{Evolution from $M_Z$ to $M_K$.
  \label{fig:eff_th}}
\end{figure}
}

A correct analysis should take into account the presence of very
different mass scales.  
At short distances, the gluonic interactions induce large logarithmic
corrections which scale as $\frac{1}{N_C}\,\log{(M_W/\mu)}$,
while at large distances they generate infrared effects of the type
$\frac{1}{N_C}\,\log{(\mu/M_\pi)}$. At
$\mu \sim 1$~GeV,  $\log{(M_W/\mu)}\sim 4$ \ and \
$\log{(\mu/M_\pi)}\sim 2$,
which breaks the $1/N_C$ expansion. 

The necessary summation of the short-distance
logarithms is performed with the operator product expansion\cite{WI:69}
(OPE) and the renormalization group. 
After integrating out all heavy scales,
one gets an effective $\Delta S=1$ \ Lagrangian, defined in the
three-flavour theory ($\mu<m_c$),\cite{GW:79}
\bel{eq:Leff}
 {\cal L}_{\mathrm{eff}}^{\Delta S=1} = - \frac{G_F}{\sqrt{2}}
 V_{ud}^{\phantom{*}}\,V^*_{us}\,  \sum_{i=1}^{10}
 C_i(\mu) \; Q_i (\mu) \, ,
 \label{eq:lag}
\ee
which is a sum of local four-fermion operators $Q_i$,
constructed with the light degrees of freedom, modulated
by Wilson coefficients $C_i(\mu)$
which are functions of the heavy masses.
The overall renormalization scale $\mu$ separates
the short- ($M>\mu$) and long- ($m<\mu$) distance contributions,
which are contained in $C_i(\mu)$ and $Q_i$, respectively.
The physical amplitudes are of course independent of $\mu$.
The Wilson coefficients have been computed at the next-to-leading
logarithmic order.\cite{buras1,ciuc1}
All gluonic corrections of $O(\alpha_s^n t^n)$ and
$O(\alpha_s^{n+1} t^n)$ are known,
where $t\equiv\ln{(M_1/M_2)}$ refers to the logarithm of any ratio of
heavy mass scales $M_1,M_2\geq\mu$.

\subsection{\chpt\ Description}

At lowest order in \chpt,
the most general effective bosonic Lagrangian
with the same $SU(3)_L\otimes SU(3)_R$ transformation properties
as the short-distance Lagrangian \eqn{eq:Leff} contains three terms,
transforming as $(8_L,1_R)$, $(27_L,1_R)$ and $(8_L,8_R)$, respectively.
Their corresponding couplings are denoted by
$g_8$, $g_{27}$ and $g_{ew}$.
At tree level, the resulting $K\to\pi\pi$ amplitudes take the form:
\beqn\label{TREE}
\cA_0 & =& -{G_F\over \sqrt{2}} \, V_{ud}^{\phantom{*}} V_{us}^*
\;\sqrt{2}\,
f_\pi\;\left\{\left ( g_8+{1\over 9}\, g_{27}\right )(M_K^2-M_\pi^2)
-{2\over 3}\, f_\pi^2\, e^2 \; g_8\; g_{ew}\right\}\; ,
\nonumber\\
\cA_2& =&  -{G_F\over \sqrt{2}} \, V_{ud}^{\phantom{*}} V_{us}^*
\; {2\over 9}
\,f_\pi\;\biggl\{ 5\, g_{27}\, (M_K^2-M_\pi^2) -3\, f_\pi^2\, e^2\; 
g_8\; g_{ew}\biggr\}
\; .
\eeqn

The isospin amplitudes $\cA_I = A_I\,\e^{i\delta_I}$ have been
computed up to next-to-leading order in 
\chpt.\cite{KA91,EMNP:00,PPS:01}
The only remaining problem is the calculation of the chiral couplings
from the short-distance Lagrangian \eqn{eq:Leff},
which requires to perform the matching between the two EFTs.
This can be easily done at $N_C\to\infty$, because
the four-quark operators factorize into currents
which have well-known chiral realizations:\cite{PPS:01}
$$
\tilde{g}_8^\infty =
\left\{ -{2\over 5}\,C_1(\mu)+{3\over 5}\,C_2(\mu)+C_4(\mu)
- 16\, L_5\, C_6(\mu)\, B(\mu)\right\}\, f_0^{K\pi}(M_\pi^2)\, ,
\hskip .1cm
$$
\bel{eq:NC_results}
\tilde{g}_{27}^\infty\, = \,
{3\over 5}\,[C_1(\mu)+C_2(\mu)]\;  f_0^{K\pi}(M_\pi^2) \, ,
\hskip 4.6cm
\ee
$$
e^2\,\left[\tilde{g}_8\,\tilde{g}_{ew}\right]^\infty_I 
 = -3\,\left\{ C_8(\mu)\, B(\mu)\,\xi_I\,\, +\,\Delta_I C(\mu)\;
{M_K^2\! -\! M_\pi^2\over 4\, f_\pi^2}\; f_0^{K\pi}(M_\pi^2)\right\}\, .
$$
The effective couplings~$\tilde{g}_I^\infty$ include the
local $O(p^4)$ \chpt\ corrections. 
They generate the factors \
$\xi_0 = 1 + 4 L_5\, M_K^2/f_\pi^2$, \
$\xi_2 = 1 + 4 L_5\, M_\pi^2/f_\pi^2$ \ and \
$f_0^{K\pi}(M_\pi^2)\approx \xi_2$, \
and introduce additional dependences on Wilson coefficients:
$\Delta_0 C(\mu) = [C_7 - C_9 + C_{10}](\mu)$, \
$\Delta_2 C(\mu) = -2\, [C_7 - C_9 - C_{10}](\mu)$.
At $N_C\to\infty$, \
$L_5^\infty = {1\over 4}\left( f_K f_\pi - f_\pi^2\right) /
\left(M_K^2-M_\pi^2\right) \approx 2.1\cdot 10^{-3}$ \
and \ $f_0^{K\pi}(M_\pi^2)\approx 1.02\, $.

The factorization of the operators $Q_i$
($i\not=6,8$) does not provide any scale dependence,
because their anomalous dimensions vanish at $N_C\to\infty$.\cite{BBG87}
To achieve a reliable expansion in powers of $1/N_C$,
one needs to go to the next order, where this physics is
captured.\cite{PR:91}
This is the reason why the study of the $\Delta I=1/2$
rule has proven to be so difficult. 
The only anomalous dimensions which survive when $N_C\to\infty$
are the ones corresponding to $Q_6$ and $Q_8$.\cite{BBG87,BG:87} 
These operators  factorize into colour-singlet
scalar and pseudoscalar currents, which are $\mu$ dependent.
This generates the factors \
$B(\mu)\equiv M_K^4/\left[ (m_s + m_q)(\mu)\, f_\pi\right]^2$
\ ($m_q\equiv m_u = m_d$),
which exactly cancel the $\mu$ dependence of
$C_{6,8}(\mu)$ at large $N_C$.\cite{BBG87,PR:91,BG:87}
Since these two operators give the leading contributions to 
Im($g_I$), the large--$N_C$ limit provides a good estimate of
the CP-violating ratio $\varepsilon'/\varepsilon$,
while Re($g_I$) gets large $1/N_C$ corrections.\cite{PR:91}

The large--$N_C$ calculation does not produce any strong phases
$\delta_I$. Those phases originate in the
final rescattering of the two pions and, therefore, are generated by
chiral loops which are of higher order in both the momentum
and $1/N_C$ expansions.
Analyticity and unitarity require a corresponding
dispersive effect in the moduli of the isospin amplitudes.
Since the S--wave phase-shift difference is very large,\cite{GM:91}
$(\delta_0 -\delta_2)(M_K^2) = 45^\circ\pm 6^\circ$,
one should expect large unitarity corrections.\cite{PaP:01}
This is confirmed by the $O(p^4)$ \chpt\ calculation,
which generates large infrared logarithms.
With the normalization
${\cal A}_I^{(R)}  = {\cal A}_I^{(R)\infty} \times\,
{\cal C}_I^{(R)}$,
the correction factors \ 
${\cal C}_I^{(R)}\equiv 1 + \Delta_L{\cal A}_I^{(R)}$ \
take the values:\cite{PPS:01}
\beqn\label{eq:onel}
{\cal C}_0^{(8)} = 1.27 \pm 0.05 +  0.46 \ i \, , \hskip .2cm
&&\hskip 5cm\no\\
{\cal C}_0^{(27)} = 2.0 \pm 0.7 + 0.46 \ i \, ,\hskip .4cm
&\qquad\qquad &
{\cal C}_2^{(27)}=  0.96 \pm 0.05-0.20 \ i\, ,
\\\
{\cal C}_0^{(ew)}=  1.27  \pm 0.05 + 0.46 \ i \, , &&
{\cal C}_2^{(ew)}=  0.50  \pm 0.24-0.20 \ i \, .
\no\eeqn
The quoted uncertainties correspond to changes of the
\chpt\ renormalization scale between 0.6 and 1~GeV.
The scale dependence is only present in the dispersive contributions
and should cancel with the corresponding
dependence of the $O(p^4)$ counterterms at the
next-to-leading order in $1/N_C$.

\subsection{The Standard Model Prediction for $\varepsilon'/\varepsilon$}

The CP-violating ratio
\bel{eq:epspdef}
{\varepsilon^\prime\over\varepsilon} =
\; e^{i\Phi}\; {\omega\over \sqrt{2}\vert\varepsilon\vert}\;\left[
{\mbox{Im}(A_2)\over\mbox{Re}(A_2)} - {\mbox{Im}(A_0)\over\mbox{Re}(A_0)}
 \right] \, ,
\qquad\quad
\Phi \approx \delta_2-\delta_0+\frac{\pi}{4}\approx 0 \, ,
\ee
constitutes a fundamental test for our understanding of 
flavour-changing phenomena.
The present experimental world average,\cite{KTEV:01,NA48:01}
\bel{eq:exp}
{\rm Re} \left(\varepsilon'/\varepsilon\right) =
(17.2 \pm 1.8) \cdot 10^{-4} \, ,
\ee
provides clear evidence for the existence of direct CP violation.

The amplitudes $\mbox{Re} (A_I)$, their ratio\
$\omega = \mbox{Re} (A_2)/\mbox{Re} (A_0) \approx 1/22$
and $\varepsilon$ are usually set to their experimentally
determined values. A theoretical calculation is then only needed
for the CP-odd quantities 
$\mbox{Im} (A_I)$, which are dominated by the
operators $Q_6$ and $Q_8$.
To a very good approximation,\cite{munich}
\bel{EPSNUM}
{\varepsilon'\over\varepsilon} \sim
\left [ B_6^{(1/2)}(1-\Omega_{IB}) - 0.4 \, B_8^{(3/2)}
 \right ]\, ,
\ee
where the factors $B_i$ parameterize the
$Q_i$ matrix elements in vacuum insertion units.
The ratio\cite{EMNP:00,PPS:01,MW:00}
\bel{eq:isospin}
\Omega_{IB} = {1\over \omega}
{\mbox{Im}(A_2)_{IB}\over \mbox{Im}(A_0)}
\,\approx\, 0.12\pm 0.05
\ee
takes into account isospin-breaking corrections, which get enhanced
by $1/\omega$.

The isospin-breaking correction 
was originally estimated to be 
$\Omega_{IB}=0.25$.\cite{BG:87,Omega}
Together with the usual ansatz $B_i\sim 1$, 
this produced a large numerical cancellation
in Eq.~\eqn{EPSNUM} leading\cite{munich,rome} 
to unphysical low values of 
$\varepsilon'/\varepsilon$ around $7\cdot 10^{-4}$.
The \chpt\ loop corrections destroy this accidental cancellation.
The final result is governed by the matrix element of the
gluonic penguin operator $Q_6$.

Taking into account all large logarithmic corrections at short
and long distances, the Standard Model prediction for
$\varepsilon'/\varepsilon$ is found to be:\cite{PPS:01}
\bel{eq:SMpred}
\mbox{\rm Re}\left(\varepsilon'/\varepsilon\right) \; =\;  
\left(1.7\pm 0.2\, {}_{-0.5}^{+0.8} \pm 0.5\right) \cdot 10^{-3}
\; =\; \left(1.7\pm 0.9\right) \cdot 10^{-3}\, ,
\ee
in excellent agreement with the measured experimental
value \eqn{eq:exp}.
The first error comes from the short-distance evaluation
of Wilson coefficients and the choice of low-energy
matching scale $\mu$.
The uncertainty coming from the strange quark mass, 
$(m_s+ m_q)(1\, \rm{GeV})=156\pm 25\, \rm{MeV}$,
is indicated by the second error.\cite{ms2002}
The most critical step is the matching between the short- and 
long-distance descriptions, which has been done
at leading order in $1/N_C$.
Since all ultraviolet
and infrared logarithms have been resummed, our educated guess for
the theoretical uncertainty associated with $1/N_C$ corrections
is $\sim 30\%$ (third error).

A better determination of the strange quark mass would allow
to reduce the uncertainty to the 30\% level.
In order to get a more accurate prediction, it would be necessary to have
a good analysis of next-to-leading $1/N_C$ corrections. This is
a very difficult task, but progress in this direction can be
expected in the next few years.\cite{PR:91,epsNLO}

\section{Summary}

The large--$N_C$ limit provides a sensible approximation to the
$N_C=3$ hadronic world. Assuming confinement,
the strong dynamics at $N_C\to\infty$ is given 
by tree diagrams with infinite sums of hadron exchanges,
which correspond to the tree approximation to some local
effective Lagrangian. Hadronic loops generate corrections
suppressed by factors of $1/N_C$.

At very low energies the hadronic EFT describing the lightest
pseudoscalar nonet is \chpt, while resonance chiral theory provides
the correct framework to incorporate the massive mesonic states.
The short-distance properties of QCD at large $N_C$ provide
strong constraints on the chiral couplings.

The expansion in powers of $1/N_C$ turns out to be a very
useful tool for quantitative non-perturbative analyses.
While there is a very successful leading-order phenomenology,
some important physical effects only appear at subleading
topologies: the $U(1)_A$ anomaly, the anomalous dimensions of
(non-penguin) four-quark operators and their associated
short-distance logarithms, the infrared \chpt\ logarithms,
the resonance widths,
etc.
Those effects can be rigorously analyzed with appropriate tools
as exemplified by the calculation of $\varepsilon'/\varepsilon$.
The control of non-logarithmic corrections at the next-to-leading
order in $1/N_C$ remains a challenge for future investigations.

\section*{Acknowledgments}
I would like to thank Richard Lebed for making possible a very stimulating
workshop. I'm also grateful to G.~Ecker and J.~Portol\'es for their
useful comments on the manuscript.
 This work has been partly supported by MCYT, Spain (Grant No.
FPA-2001-3031) and by the European Union TMR network
{\it EURODAPHNE} (Contract No. ERBFMX-CT98-0169).


\end{document}